\documentclass[10pt]{iopart}

\usepackage[utf8]{inputenc}
\usepackage[T1]{fontenc}
\usepackage{float} 
\usepackage{color}  
\usepackage{graphicx}
\usepackage{amssymb}
\usepackage{bbm}
\usepackage{indentfirst}
\usepackage{dcolumn}
\usepackage{soul}
\usepackage{mathrsfs}
\usepackage{xcolor}

\newcommand{\ch}[1]{\textcolor{red}{#1}}

\begin{document}

\title{Spin waves in skyrmionic structures with various topological charges}

\author{Levente R\'{o}zsa$^{1}$, Markus Wei{\ss}enhofer$^{1}$ and Ulrich Nowak$^{1}$}
\address{$^{1}$ Department of Physics, University of Konstanz, D-78457 Konstanz, Germany}
\ead{levente.rozsa@uni-konstanz.de}
\date{\today}

\begin{abstract}

Equilibrium properties and localised magnon excitations are investigated in topologically distinct skyrmionic textures. The observed shape of the structures and their orientation on the lattice is explained based on their vorticities and the symmetry of the crystal. The transformation between different textures and their annihilation as a function of magnetic field is understood based on the energy differences between them. 
%their relative energies.
%, with the topological charges playing a lesser role. 
The angular momentum spin-wave eigenmodes characteristic of cylindrically symmetric structures are combined in the distorted spin configurations, leading to avoided crossings in the magnon spectrum. The susceptibility of the skyrmionic textures to homogeneous external fields is calculated, revealing that a high number of modes become detectable due to the hybridization between the angular momentum eigenmodes. These findings should contribute to the observation of spin waves in distorted skyrmionic structures via experiments and numerical simulations, widening the range of their possible applications in magnonic devices.

\end{abstract}

%\keywords{skyrmions, magnons, linear response theory}

\maketitle
\ioptwocol

\section{Introduction}

Isolated magnetic skyrmions are localised spin textures embedded in homogeneous magnetic phases~\cite{Bogdanov,Nagaosa}. Since they may easily be manipulated and detected by magnetic, electrical and optical means, they continue to attract considerable research interest as possible elements in conventional~\cite{Fert} and unconventional~\cite{Zazvorka,Jibiki} computing architectures.

Utilizing the small-amplitude spin-wave excitations of skyrmions enables their more energy-efficient application in magnonic devices, particularly in insulators where possibilities for electrically moving the spin configurations are limited~\cite{Garst}. The characteristic magnon modes of both the skyrmion lattice phase~\cite{Petrova,Zang,Mochizuki} and isolated magnetic skyrmions~\cite{Lin,Schutte} have been determined based on theoretical calculations and numerical simulations, and also observed experimentally~\cite{Schwarze,Onose,Satywali}. Skyrmion lattices provide a realization of topological magnon insulators~\cite{Shindou,Zhang}, and topological edge modes have been identified in them~\cite{Roldan-Molina,Diaz}. The spin wave modes have been determined in skyrmion strings in bulk magnets~\cite{Seki}, in so-called $k\pi$ skyrmions~\cite{Rozsa3} and in skyrmions in antiferromagnets~\cite{Kravchuk2} as well.

The most widely studied mechanism for the stabilisation of skyrmions is based on the Dzyaloshinsky--Moriya interaction~\cite{Dzyaloshinsky,Moriya} present in noncentrosymmetric magnets~\cite{Bogdanov}. By competing with the isotropic Heisenberg, uniaxial anisotropy and Zeeman terms preferring a collinear alignment of the spins, this chiral interaction gives rise to cylindrically symmetric isolated skyrmions. This symmetry considerably simplifies the numerical treatment of their equilibrium profile~\cite{Bogdanov} and their spin wave eigenmodes~\cite{Bogdanov4,Schutte,Kravchuk2}, which only have to be determined along the radial direction from the center of the structure. The angular dependence of the magnon modes may be classified according to the angular momentum eigenvalues of the rotation operator around the axis of the skyrmion. The possible experimental detection of the spin wave modes by applying a spatially homogeneous external excitation is limited by the selection rules for the angular momentum quantum numbers~\cite{Schutte}. While theoretical calculations managed to identify a high number of excitations both in skyrmion lattices and isolated skyrmions~\cite{Schutte,Kravchuk2,Rozsa3}, their observation in experiments~\cite{Schwarze,Buttner2} and numerical simulations~\cite{Mochizuki} is restricted to breathing modes which may be excited by out-of-plane fields, and to gyration modes reacting to in-plane fields, since these possess a finite oscillating magnetic dipolar moment that couples to the external field.

Notwithstanding the success of the micromagnetic model describing chiral skyrmions, such localised spin structures often deviate from the cylindrically symmetric shape. In experimental realizations this may be caused by the inhomogeneities introduced during the growth process~\cite{Jiang}, by strain relief due to the lattice mismatch in epitaxially grown systems~\cite{Hsu} or by intermixing and defects~\cite{Meyer}. If the symmetry of the system is sufficiently low, the Dzyaloshinsky--Moriya interaction~\cite{Hoffmann} and thus the whole micromagnetic model will be modified such that cylindrically symmetric solutions are no longer permitted. The shape of the skyrmions may also be deformed intentionally by applying mechanical strain~\cite{Shibata,Sukhanov} or magnetic fields tilted with respect to the anisotropy axis~\cite{Ikka,Koide,Kuchkin,Gobel}.
% The cross-section of isolated skyrmions stabilised in the conical phase of chiral magnets deviates from the circular shape.

Another mechanism for the stabilisation of skyrmions involves the competition between ferromagnetic and antiferromagnetic Heisenberg exchange interactions with different neighbours in the crystal, preferring the formation of modulated spin structures~\cite{Okubo,Leonov,Kurumaji}. Since these frustrated isotropic interactions do not select a preferred rotational sense for the textures, skyrmions and antiskyrmions with the in-plane components of the spins rotating in opposite directions may be stabilised simultaneously. It was demonstrated in \cite{Rozsa2} that additionally considering the Dzyaloshinsky--Moriya interaction selects a single type of skyrmionic structure where the rotational sense of the spins agrees with the direction preferred by the chiral interaction in the whole object, and distorts the shape of all other skyrmionic textures by extending and shrinking the regions with favoured and unfavoured rotational senses, respectively. It has been shown recently that such distorted spin configurations are also stable in chiral magnets in the absence of frustrated isotropic interactions~\cite{Kuchkin2}. The asymmetric shape and the alignment of the skyrmionic structures along preferential directions on the lattice has been demonstrated to have a profound effect on their current-driven dynamics~\cite{Ritzmann,Weissenhofer,Chen} and thermal diffusion~\cite{Kerber}.

Despite the prevalence of not cylindrically symmetric skyrmions in experimental systems and in theoretical models, their spin wave modes have not been explored in detail so far. In Ref.~\cite{Ikka}, it was demonstrated by numerical simulations that besides the previously observed clockwise and counterclockwise gyration modes and the breathing mode of the skyrmion lattice stabilised by the Dzyaloshinsky--Moriya interaction, an additional excitation mode may be observed when the stabilizing field is not perpendicular to the surface. Based on the terminology developed for the cylindrically symmetric case and on the investigation of its spatial profile, this additional mode was classified as another clockwise gyration mode. In numerical simulations performed for skyrmions confined in nanodots in ferromagnetic~\cite{Kim} and ferrimagnetic~\cite{Lonsky} materials, more pronounced deviations from the radially symmetric profiles of the breathing modes have been observed for out-of-plane excitations.

In this work, the spin wave modes in distorted skyrmionic spin structures are investigated systematically, focusing on isolated spin textures with various topological charges stabilised by the competition between frustrated exchange interactions and the Dzyaloshinsky--Moriya interaction. It is demonstrated how the distortion of the structure transforms the angular momentum eigenmodes to generic modes which cannot be described by a single angular momentum quantum number. Hybridizations of the modes and avoided crossings in the magnon spectrum are discussed. Since most of the eigenmodes contain a breathing character, it is discussed how the localised spin waves may be detected by applying an out-of-plane oscillating field.

The paper is organized as follows. In section~\ref{sec2}, the atomistic spin model used for the numerical calculations is introduced, using the spin model parameters determined in \cite{Rozsa}. 
% based on the electronic structure.
The static properties of skyrmions in frustrated magnets with the Dzyaloshinsky--Moriya interaction are discussed in general in section~\ref{sec2b}, and applied to the considered system in section~\ref{sec2b3}. The spin wave eigenmodes of the structures are characterized in section~\ref{sec2b2}, and a formula describing their susceptibility to oscillating external excitations is derived in section~\ref{sec2d}. A summary and discussion of the results concludes the work in section~\ref{sec4}.

\section{Methods\label{sec2}}

%\subsection{Atomistic spin model and spin wave spectrum\label{sec2a}}

The skyrmionic structures are described within the atomistic spin model
\begin{eqnarray}
H=\frac{1}{2}\sum_{i \ne j}\bi{S}_{i}\mathcal{J}_{ij} \bi{S}_{j}+\sum_{i}\bi{S}_{i}\mathcal{K} \bi{S}_{i}-\sum_{i}\mu_{\textrm{s}}\bi{S}_{i}\bi{B},\label{eqn1}
\end{eqnarray}
where the $\bi{S}_{i}$ denote unit vectors describing classical spins or magnetic moments on the sites of a single-layer triangular lattice. The magnetic moments of magnitude $\mu_{\textrm{s}}$ are coupled to the external magnetic field $\bi{B}$ and to each other through the exchange interaction tensors $\mathcal{J}_{ij}$ and the on-site anisotropy tensor $\mathcal{K}$. The $\mathcal{J}_{ij}$ coupling tensors include the isotropic Heisenberg exchange interactions,
\begin{eqnarray}
J_{ij}=\frac{1}{3}\textrm{Tr}\mathcal{J}_{ij},\label{eqn2}
\end{eqnarray}
the antisymmetric Dzyaloshinsky--Moriya interactions,
\begin{eqnarray}
\bi{D}_{ij}\left(\bi{S}_{i}\times\bi{S}_{j}\right)=\frac{1}{2}\bi{S}_{i}\left(\mathcal{J}_{ij}-\mathcal{J}^{T}_{ij}\right)\bi{S}_{j},\label{eqn3}
\end{eqnarray}
and a symmetric traceless part contributing to the anisotropy,
\begin{eqnarray}
\mathcal{J}^{\textrm{S}}_{ij}=\frac{1}{2}\left(\mathcal{J}_{ij}+\mathcal{J}^{T}_{ij}\right)-J_{ij}\bi{I}.\label{eqn4}
\end{eqnarray}
The on-site anisotropy tensor $\mathcal{K}$ is reduced to a uniaxial out-of-plane anisotropy constant $K^{zz}$, since the system possesses a $C_{3\textrm{v}}$ symmetry. The magnetic field is oriented along the out-of-plane direction in all considered cases.

The selected parameters describe the interactions between the Fe magnetic moments in the (Pt$_{0.95}$Ir$_{0.05}$)/Fe/Pd(111) ultrathin film system. The values of the coupling coefficients were determined based on the screened Korringa--Kohn--Rostoker method~\cite{Szunyogh} in \cite{Rozsa}, with the numerical values reported in \cite{Rozsa2,Zazvorka}. The exchange tensors were calculated up to a distance of $8$ lattice constants between the spins, showcasing an oscillatory decay as in the Ruderman--Kittel--Kasuya--Yosida mechanism. The interplay between the frustrated Heisenberg interactions with the Dzyaloshinsky--Moriya interaction in the material stabilises magnetic skyrmionic structures with diverse shapes and values of the topological charge~\cite{Rozsa2}.

The dynamics of the magnetic moments is described based on the Landau--Lifshitz--Gilbert equation~\cite{Landau,Gilbert},
\begin{eqnarray}
\frac{\textrm{d}\bi{S}_{i}}{\textrm{d}t}=-\gamma' \bi{S}_{i} \times \bi{B}_{i}^{\textrm{eff}} - \gamma' \alpha\bi{S}_{i} \times\left(\bi{S}_{i} \times \bi{B}_{i}^{\textrm{eff}}\right),\label{eqn5}
\end{eqnarray}
with $\alpha$ the Gilbert damping parameter, $\gamma'=\frac{ge}{2m\left(1+\alpha^{2}\right)}$ the modified gyromagnetic ratio of the electron (with $e$ denoting the elementary charge, $g$ the spin $g$-factor and $m$ the mass of the electron), and $\bi{B}_{i}^{\textrm{eff}}=-\frac{1}{\mu_{\textrm{s}}}\frac{\partial H}{\partial \bi{S}_{i}}$ the effective field acting on $\bi{S}_{i}$. The equilibrium configurations were determined by numerically solving the Landau--Lifshitz--Gilbert equation, which leads to the minimization of the torque $\bi{S}_{i} \times \bi{B}_{i}^{\textrm{eff}}$. Close to equilibrium the relaxation was sped up by rotating the direction of the spins to become parallel with the effective field direction in each step. The configuration was considered converged when the torque at each lattice site became smaller than $10^{-8}\,\textrm{mRy}/\mu_{\textrm{B}}$, given in magnetic field units.

After finding the equilibrium spin configuration, the Hamiltonian in \eref{eqn1} was expanded up to second order in small transversal deviations from this state,
\begin{eqnarray}
H\approx H_{0}+\frac{1}{2}\left(\tilde{\bi{S}}^{\perp}\right)^{T}\bi{H}_{\textrm{SW}}\tilde{\bi{S}}^{\perp},\label{eqn6}
\end{eqnarray}
where $\tilde{\bi{S}}^{\perp}$ is a vector of length $2N$, containing the $x$ and $y$ components of the spin vectors in the local frame of reference where the $z$ axis is pointing along the equilibrium spin direction at each of the $N$ lattice sites. Substituting \eref{eqn6} into \eref{eqn5} yields the linearised Landau--Lifshitz--Gilbert equation,
\begin{eqnarray}
\partial_{t}\tilde{\bi{S}}^{\perp}=\frac{\gamma'}{\mu_{\textrm{s}}}\left(-\textrm{i}\bi{\sigma}^{y}-\alpha\right)\bi{H}_{\textrm{SW}}\tilde{\bi{S}}^{\perp}=\bi{D}_{\textrm{SW}}\tilde{\bi{S}}^{\perp},\label{eqn7}
\end{eqnarray}
which reduces to an eigenvalue equation after performing Fourier transformation in time. $\bi{\sigma}^{y}$ in \eref{eqn7} denotes the Pauli matrix acting on the subspace of the $x$ and $y$ components in the local frame of reference. The eigenvalues $-\textrm{i}\omega_{k}$ and the right eigenvectors $\bi{r}^{k}$ correspond to the spin-wave frequencies and modes. More details on the structures of the matrices $\bi{H}_{\textrm{SW}}$ and $\bi{D}_{\textrm{SW}}$ are given in, e.g., \cite{Rozsa3}. The calculations were performed on a $128\times128$ lattice, which was significantly larger than the size of the isolated skyrmionic structures at all considered field values {\ch{in order to avoid interactions between repeated images due to the periodic boundary conditions}}.

\section{Results}

\subsection{Classification of skyrmionic structures in the continuum limit\label{sec2b}}

The skyrmionic structures and their eigenmodes will be described here in the micromagnetic limit, where the continuous rotational and translational symmetries simplify this characterization compared to the lattice model with discrete symmetries. In the regime where the size of the modulated spin structures is significantly larger than the lattice constant, the atomistic model given in \eref{eqn1} may be well approximated by the free-energy functional
\begin{eqnarray}
\mathscr{H}=&&\int \Big[-\mathscr{J}_{1}\left(\bi{\nabla}\bi{S}\right)^{2}+\mathscr{J}_{2}\left(\bi{\nabla}^{2}\bi{S}\right)^{2}+\mathscr{D}w_{\textrm{DM}}\left(\bi{S}\right)\nonumber
\\
&&-\mathscr{K}\left(S^{z}\right)^{2}-\mathscr{M}\bi{S}\bi{B}\Big]\textrm{d}^{2}\bi{r},\label{eqn8}
\end{eqnarray}
where $\bi{S}\left(\bi{r}\right)$ is the spin vector field of unit length defined in the two-dimensional plane. The parameters $\mathscr{J}_{1},\mathscr{J}_{2}>0$ describe the isotropic exchange interactions. By performing Fourier transformation in space ($\bi{\nabla}\rightarrow\textrm{i}\bi{k}$), %it becomes clear that the isotropic terms themselves prefer a modulated magnetic order with a finite wave vector,
the isotropic terms assume the wave-vector dependence $-\mathscr{J}_{1}\bi{k}^{2}+\mathscr{J}_{2}\left(\bi{k}^{2}\right)^{2}$, clearly preferring a modulated magnetic order with a finite wave vector $\bi{k}_{0}^{2}=\mathscr{J}_{1}/\left(2\mathscr{J}_{2}\right)$. In the atomistic description, this is caused by the presence of both ferromagnetic and antiferromagnetic Heisenberg interactions between pairs of spins located at different distances. The Dzyaloshinsky--Moriya interaction of strength $\mathscr{D}$ is described by the linear Lifshitz invariant
\begin{eqnarray}
w_{\textrm{DM}}\left(\bi{S}\right)=S^{z}\partial_{x}S^{x}-S^{x}\partial_{x}S^{z}+S^{z}\partial_{y}S^{y}-S^{y}\partial_{y}S^{z},\label{eqn9}
\end{eqnarray}
respecting the $C_{3\textrm{v}}$ symmetry of the lattice model. The coefficient $\mathscr{K}$ stands for the uniaxial anisotropy, while $\mathscr{M}$ is the magnetisation.

The spin field may be rewritten in spherical coordinates,
\begin{eqnarray}
\bi{S}=\left[\begin{array}{c}\sin\Theta\cos\Phi \\ \sin\Theta\sin\Phi \\ \cos\Theta\end{array}\right].\label{eqn9a}
\end{eqnarray}
For the description of skyrmionic structures it is advantageous to use polar coordinates in the plane, $\bi{r}=\left(r,\varphi\right)$. 
% In the most general case,
In these coordinates, the free energy in \eref{eqn8} may be expressed as
\begin{eqnarray}
\mathscr{H}=\int h\left(\partial^{n_{\Theta r}}_{r}\partial^{n_{\Theta \varphi}}_{\varphi}\Theta,\partial^{n_{\Phi r}}_{r}\partial^{n_{\Phi \varphi}}_{\varphi}\Phi,r,\varphi\right)\textrm{d}^{2}\bi{r},\label{eqn10}
\end{eqnarray}
where $n_{\Theta r},n_{\Theta \varphi},n_{\Phi r},n_{\Phi \varphi}\in\mathbb{N}$ denote the orders of partial derivatives with respect to $r$ and $\varphi$. Instead of writing out the free-energy density $h$ explicitly, here we will discuss only its symmetry properties. 
%, which description may be generalized if \eref{eqn8} is extended by additional interaction terms, e.g., by the dipolar interaction, exchange anisotropy or higher-order spin couplings.
Due to the translational invariance, the explicit dependence of $h$ on $r$ and $\varphi$ may be dropped. Assuming the external field $\bi{B}$ is oriented along the out-of-plane direction, the free energy in \eref{eqn8} is also invariant under rotation around the $z$ axis by an arbitrary angle $\chi$, which transforms the fields and variables as
\begin{eqnarray}
\Theta\rightarrow\Theta,\:\Phi\rightarrow\Phi+\chi,\:r\rightarrow r,\:\varphi\rightarrow\varphi+\chi.\label{eqn11}
\end{eqnarray}
This implies that $h$ may only depend on $\Phi$ in the form $\Phi-\varphi$, or through the derivatives of $\Phi$.

Due to the rotational invariance, we will look for cylindrically symmetric equilibrium states of \eref{eqn8},
\begin{eqnarray}
\Theta_{0}\left(r,\varphi\right)&=&\Theta_{0}\left(r\right),\label{eqn12}
\\
\Phi_{0}\left(r,\varphi\right)&=&M\varphi+\varphi_{0},\label{eqn13}
\end{eqnarray}
where the vorticity $M$ describes the number and the direction of $2\pi$ rotations of the in-plane spin components on a contour encircling the origin, and the helicity $\varphi_{0}$ determines the phase of this rotation~\cite{Nagaosa}. Configurations with $M>0$ are known as skyrmions, while structures with $M<0$ are antiskyrmions. The topological charge of the skyrmionic structures is given by $Q=-\textrm{sgn}\left(B^{z}\right)M$~\cite{Leonov,Rozsa2}.

The isotropic terms, the anisotropy, and the Zeeman term in \eref{eqn8} do not depend explicitly on $\Phi$. They are also invariant under spatial inversion,
\begin{eqnarray}
\Theta\rightarrow\Theta,\:\Phi\rightarrow\Phi,\:r\rightarrow r,\:\varphi\rightarrow -\varphi;\label{eqn14}
\end{eqnarray}
note that the spin field, being an axial vector, is not reversed under spatial inversion. This implies that only even derivatives of $\Phi$ with respect to the angle $\varphi$ enter $h$, i.e., $n_{\Phi \varphi}=2,4,\dots$. From this follows that the helicity $\varphi_{0}$ is a free parameter, and that the free energy only depends on the square of the vorticity $M^{2}$; consequently, skyrmions and antiskyrmions are energetically equivalent in this model~\cite{Leonov}. The skyrmionic structures are stabilised by the frustrated Heisenberg interactions, $\mathscr{J}_{1},\mathscr{J}_{2}>0$.

In \eref{eqn8}, only the Dzyaloshinsky--Moriya interaction introduces a dependence on $\Phi-\varphi$, which reduces to the constant $\varphi_{0}$ when $M=1$. Skyrmionic structures with other vorticities may still be stabilised in this description, but they will become distorted from the cylindrically symmetric shape of \eref{eqn12} and \eref{eqn13} as they try to minimize the energy contribution from the Dzyaloshinsky--Moriya interaction~\cite{Rozsa2,Kuchkin2}. In single skyrmions with $M=1$, the helicity also becomes locked to $\varphi_{0}=0$ or $\pi$ depending on the sign of $\mathscr{D}$, while in other skyrmionic structures it is adapted to the distorted shape. The locking of the helicity and the lifting of the degeneracy between positive and negative vorticities also occurs if long-range dipolar interactions or local compass anisotropy terms~\cite{Nussinov} are taken into account, the latter being included in the symmetric traceless part of the interaction tensors in \eref{eqn4}.

\subsection{Skyrmionic structures in equilibrium\label{sec2b3}}

\begin{figure*}
\centering
\includegraphics[width=2.0\columnwidth]{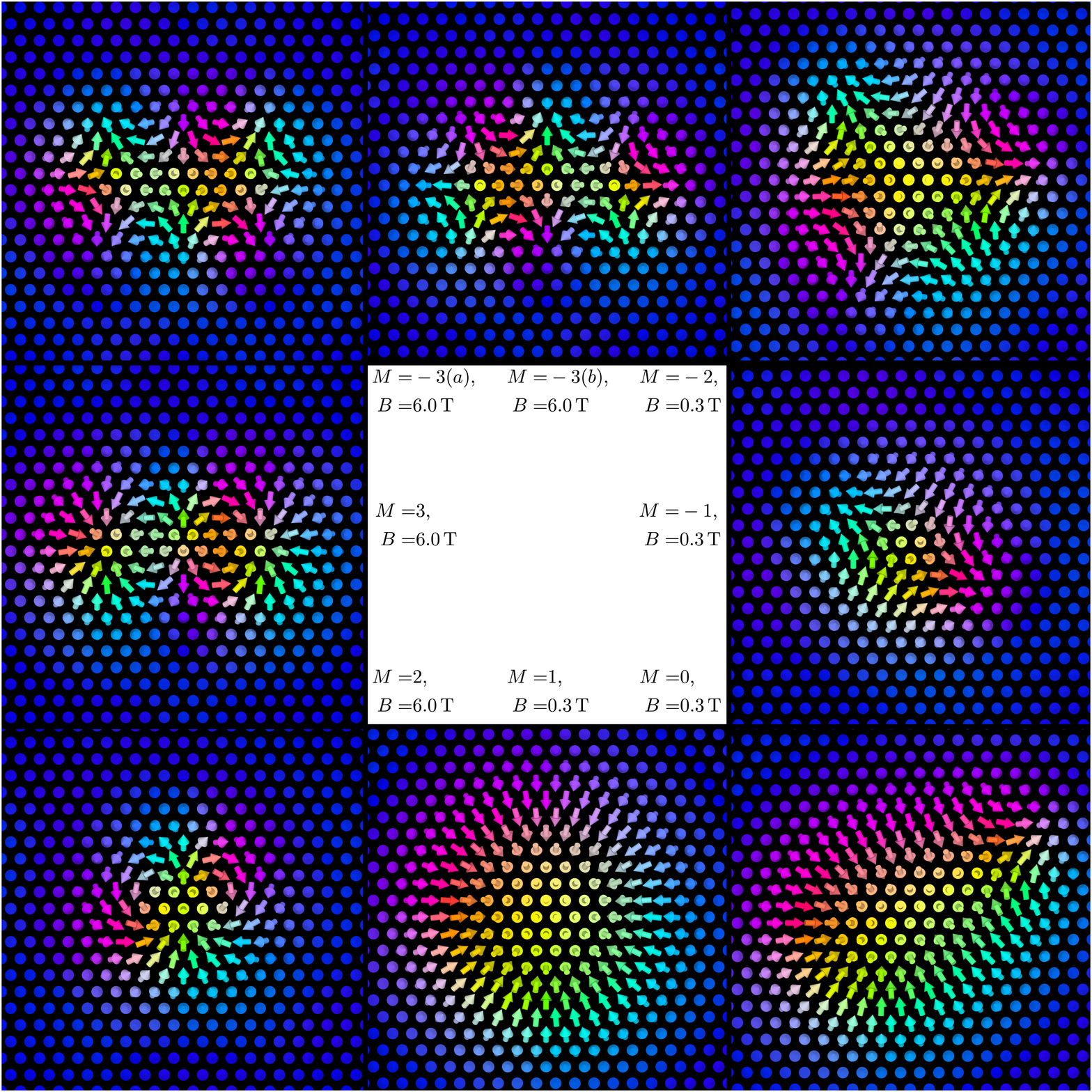}
\caption{Various types of skyrmionic structures in the (Pt$_{0.95}$Ir$_{0.05}$)/Fe/Pd(111) system, with vorticities ranging from $M=-3$ to $M=3$. The magnetic field $B$ was oriented outwards from the surface, taking the values indicated in the figure.\label{fig2}}
\end{figure*}

Returning to the atomistic model described in section~\ref{sec2}, several different types of isolated skyrmionic configurations found in the considered system are illustrated in figure~\ref{fig2}. It was discussed in \cite{Rozsa2} 
%as the structures become distorted by the Dzyaloshinsky--Moriya interaction,
that structures with vorticity $M\neq 1$ display a $C_{\left|1-M\right|}$ symmetry, as can be seen for $M=-2,-1,0,2$, and $3$ in the figure. This also holds in the continuum model, and can be deduced from \eref{eqn11} and \eref{eqn13}, since rotating the spin structure by $2\pi/\left|1-M\right|$ leaves the helicity invariant, transforming the skyrmionic configuration into itself.

The $C_{3\textrm{v}}$ symmetry of the underlying atomic lattice restricts the possible choice of equilibrium configurations further compared to the cylindrically symmetric free-energy functional in \eref{eqn8}.
% This is most pronounced for the two antiskyrmions with vorticity $M=-3$ in figure~\ref{fig2}.
First, the helicity becomes locked to the lattice, creating preferred orientations for the skyrmionic structures~\cite{Rozsa2}. These orientations are robust against small perturbations such as weak driving currents or thermal fluctuations at low temperature~\cite{Weissenhofer,Rozsa3}. However, the $C_{4}$ symmetry of the $M=-3$ antiskyrmion is incompatible with that of the atomic lattice, and is further reduced to $C_{2}$. The $M=-3\,(a)$ and $M=-3\,(b)$ structures in figure~\ref{fig2} differ in helicity by $\pi$, as can be read down from the color-coding of the in-plane spin components. Without taking into account the lattice structure, these objects could be transformed into each other by a $\pi/4$ rotation, but here they are no longer energetically degenerate.
%, and their localised spin wave excitations also differ.
% While these would be equivalent for a four-fold rotational symmetry, here they are no longer energetically degenerate, and their spin wave excitations also differ.

It is worth mentioning that the vorticity locking plays an important role in the stabilisation of the chimera skyrmion with $M=0$. This structure typically emerges as an unstable intermediate state during the creation or annihilation of normal skyrmions stabilised by the Dzyaloshinsky--Moriya interaction~\cite{Buttner,Wieser,Muckel}. It can be observed as a metastable spin structure if 
% the Heisenberg exchange interactions are frustrated, and simultaneously 
the Dzyaloshinsky--Moriya interaction locks the in-plane spin components sufficiently strongly that it can prevent the unwinding of this topologically trivial structure.
% It was not possible to stabilise the chimera skyrmion for scaling parameter values $\varepsilon\le 0.8$ in Eqs.~\eref{eqn16a} and \eref{eqn16b}.

\begin{table}
\caption{Lower and upper critical field values for the isolated skyrmionic objects in figure~\ref{fig2}. The type of instability at the critical fields is also listed in the table.\label{table1}}
\begin{indented}
\item[]
\begin{tabular}{rrrrr}
\br
         $M$ & $B_{\textrm{c,low}}$ (T) & instability type & $B_{\textrm{c,high}}$ (T) & instability type \\
\mr
        -3 $(a)$ &       0.18 & strip-out &      17.19 & to $M=-2$ \\

        -3 $(b)$ &       3.22 & to $M=-3\,(a)$ &      16.71 & to $M=-2$ \\

        -2 &       0.16 & strip-out &      24.38 &   collapse \\

        -1 &     -0.149 &      burst &      15.26 &   collapse \\

         0 &       0.19 & strip-out &      0.611 &   collapse \\

         1 &       0.08 & strip-out &      19.32 &   collapse \\

         2 &       2.17 & to two $M=1$ &      20.29 &   collapse \\

         3 &       0.97 & to three $M=1$ &      19.92 &   collapse \\
\br
\end{tabular}
\end{indented}  
\end{table}

Results concerning the stability of the skyrmionic structures as a function of external magnetic field are summarized in Table~\ref{table1}. {\ch{The types of the instabilities were identified by the numerical relaxation of the equilibrium spin configuration, and it was confirmed that all instabilities are accompanied by one of the eigenvalues of the spin-wave Hamiltonian $\bi{H}_{\textrm{SW}}$ in \eref{eqn6} approaching zero at the critical field value.}} The chimera skyrmion is stable in the by far narrowest field regime, reflecting the delicate conditions required for its stabilisation discussed above. While the other structures collapse only at very high field values where their size becomes comparable to the lattice constant, the chimera skyrmion can become unwinded already at $B=0.611\,\textrm{T}$ at a significantly larger size. Such a collapse mechanism is typical for large magnetic bubble domains stabilised only by the dipolar interaction and can also be interpreted in the continuum description~\cite{Kiselev}, while skyrmions stabilised by the Dzyaloshinsky--Moriya interaction are allowed to shrink to arbitrarily small sizes without collapsing when lattice discretisation is neglected.

The normal skyrmion with $M=1$ collapses into the collinear background at high field values, and elongates elliptically at low fields. This strip-out instability is typical for skyrmions stabilised by the Dzyaloshinsky--Moriya interaction if the system has a spin spiral ground state~\cite{Bogdanov3,Schutte}. In the present model, the spin spiral ground state transforms directly into a field-polarized state at $B=0.21\,\textrm{T}$, with no region between where a skyrmion lattice state would be energetically the most preferable~\cite{Rozsa2}. Similar instabilities can be observed for the chimera skyrmion and the $M=-3\,(a)$ and $M=-2$ antiskyrmions, the latter elongating along the three directions where the rotational sense of the spins is unfavoured by the Dzyaloshinsky--Moriya interaction; see the in-plane spin components pointing outwards from the center of the structure in figure~\ref{fig2}. It should be noted that while the $M=1$ skyrmion is the only configuration preferred by the Dzyaloshinsky--Moriya interaction, it is not the most robust one against adjustments in the external magnetic field. The $M=-1$ antiskyrmion remains stable down to negative field values -- i.e., with $\bi{B}$ oriented along the magnetisation in its core --, where it undergoes a burst instability, increasing in size more or less isotropically. This method of destruction is typical for isolated skyrmions in systems where the ground state is ferromagnetic~\cite{Bogdanov4}, and was also calculated to occur at positive field values for skyrmions with multiple spin rotations along the radial directions~\cite{Rozsa3}.

{\ch{We define the energy of the skyrmionic objects with respect to the field-polarized background normalized to unit topological charge or vorticity, $E_{\textrm{norm}}\left(M\right)=\left(E\left(M\right)-E_{\textrm{FP}}\right)/\left|M\right|$. Comparing $E_{\textrm{norm}}\left(M\right)$ between different structures reveals the possible transition mechanisms between them. The structures with vorticities $M=-2,2$, and $3$ are stable up to higher field values than the $M=1$ skyrmion. Not only the energy barrier separating them from the field-polarized background becomes higher than that of the $M=1$ skyrmion in this limit, which defines at what magnetic field they will collapse, but their $E_{\textrm{norm}}\left(M\right)$ value also becomes lower.
}}
% Not only the energy barrier separating them from the field-polarized background becomes higher than that of the $M=1$ skyrmion in this limit, which defines at what magnetic field they will collapse, but also the energy difference between the skyrmionic object and the field-polarized background normalized to unit topological charge, characterizing the binding energy of the structures.%~\cite{Rozsa2}.

In certain cases, increasing or decreasing the external magnetic field transforms one type of localised configuration into another. The $M=2$ skyrmion splits into two $M=1$ skyrmions at $B=2.17\,\textrm{T}$, which is possible because {\ch{$E_{\textrm{norm}}\left(M=2\right)>E_{\textrm{norm}}\left(M=1\right)$ in this field range.}} 
% its binding energy is positive in this field range. 
Note that this critical field value is higher than what was reported from Ref.~\cite{Rozsa2}, where less stringent convergence criteria were applied on the spin configurations. The $M=3$ skyrmion is destroyed at an even lower field value, where it can only split into three $M=1$ skyrmions since the $M=2$ skyrmion is no longer stable. The $M=-3\,(b)$ antiskyrmion transforms into the $M=-3\,(a)$ structure at low field, by only changing its vorticity. In this regime, {\ch{$E_{\textrm{norm}}\left(M=-3(a,b)\right)<E_{\textrm{norm}}\left(M=-2\right)<E_{\textrm{norm}}\left(M=-1\right)$ holds,}} 
%the binding energy of the $M=-2$ object is negative compared to the $M=-1$ antiskyrmion, and that of the $M=-3$ structures is even lower, 
which prohibits the splitting mechanisms observed for the skyrmions. At high field values, both types of the $M=-3$ antiskyrmion transform into the $M=-2$ structure, which transition does not conserve the topological charge. However, note that all possible transition routes where the topological charge would remain constant are energetically prohibited at this field value, since the $M=-1$ antiskyrmion is no longer stable, and, e.g., two $M=-2$ antiskyrmion and a single $M=1$ skyrmion would have a higher total energy. These results demonstrate that while the topological charge in the lattice model often gives strong hints concerning the stability of the structures and the possible transition routes between them, it cannot be treated as a strictly conserved quantity.% like in the continuum limit.

\subsection{Spin wave eigenmodes of skyrmionic structures\label{sec2b2}}

\begin{figure}
\includegraphics[width=\columnwidth]{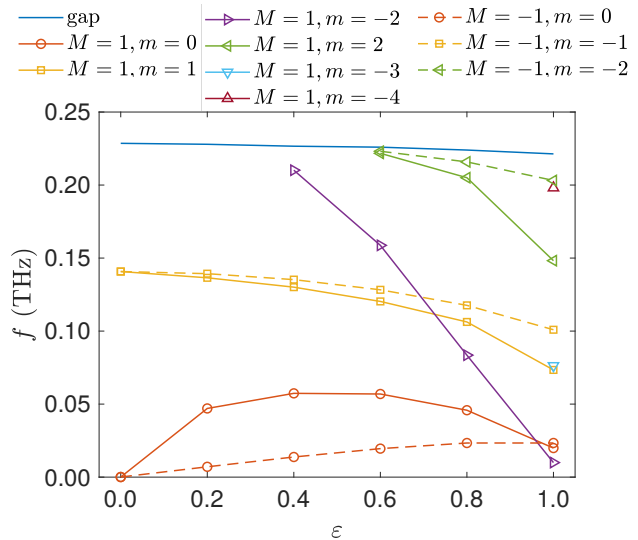}
\caption{Dependence of the spin wave frequencies on the scaling coefficient $\varepsilon$ (see \eref{eqn16a} and \eref{eqn16b}), for the skyrmion ($M=1$) and the antiskyrmion ($M=-1$). The magnetic field of strength $B=0.23\,\textrm{T}$ was oriented outwards from the surface.\label{fig1}}
\end{figure}

We will begin the classification of the low-amplitude excitations in the continuum limit, where the higher symmetry of the free energy in \eref{eqn8} and of the spin structures simplifies the description via multiple conserved quantities. After finding the equilibrium spin configuration $\Theta_{0},\Phi_{0}$ in \eref{eqn12} and \eref{eqn13}, the spin wave frequencies and modes in the continuum description may be identified analogously to the atomistic model. For cylindrically symmetric structures, the spin waves may be sought in the form
\begin{eqnarray}
\Theta_{nm}\left(r,\varphi\right)&=&\Theta_{0}\left(r\right)+\delta\Theta_{nm}\left(r\right)\textrm{e}^{\textrm{i}m\varphi},\label{eqn15}
\\
\Phi_{nm}\left(r,\varphi\right)&=&M\varphi+\varphi_{0}+\delta\Phi_{nm}\left(r\right)\textrm{e}^{\textrm{i}m\varphi},\label{eqn16}
\end{eqnarray}
where they are characterized by the main quantum number $n$ and the angular momentum quantum number $m$. For skyrmionic structures consisting of a single closed domain wall, including all configurations shown in figure~\ref{fig1}, typically only the $n=1$ modes are localised at the object, and the $n$ quantum number will be dropped in the following. As mentioned in section~\ref{sec2b}, in the absence of the Dzyaloshinsky--Moriya interaction in \eref{eqn8}, only even derivatives of $\varphi$ appear in the free energy. Since $\partial^{2}_{\varphi}$ is replaced by $-M^{2}$, $-m^{2}$ or $-mM$ when using the trial functions in \eref{eqn15} and \eref{eqn16}, this means that for an eigenmode $m$ of a skyrmion with $M>0$, the equivalent antiskyrmion with $M<0$ will have a spin wave eigenmode with the same frequency and radial spatial dependence, but the angular momentum $-m$. The sign of the angular momentum quantum number determines the rotational direction of the skyrmionic structure in the given spin wave mode~\cite{Rozsa2}, implying that skyrmions and antiskyrmions in a frustrated Heisenberg magnet rotate in opposite directions with the same frequencies.

To illustrate how the spin wave eigenmodes of skyrmionic structures in an isotropic frustrated system are modified by the Dzyaloshinsky--Moriya interaction and two-site anisotropy terms, the exchange and anisotropy tensors in \eref{eqn1} were scaled using the parameter $\varepsilon$ as
\begin{eqnarray}
\mathcal{J}_{ij}^{\textrm{scaled}}\left(\varepsilon\right)&=&J_{ij}\bi{I}+\varepsilon\left(\mathcal{J}_{ij}-J_{ij}\bi{I}\right),\label{eqn16a}
\\
\mathcal{K}^{\textrm{scaled}}\left(\varepsilon\right)&=&\mathcal{K}+\left(1-\varepsilon\right)\frac{1}{2}\sum_{j\neq i}\mathcal{J}^{\textrm{S}}_{ij}.\label{eqn16b}
\end{eqnarray}
For $\varepsilon=1$, the interactions obtained from \textit{ab initio} calculations for the (Pt$_{0.95}$Ir$_{0.05}$)/Fe/Pd(111) system are recovered. For $\varepsilon=0$, only the isotropic parts of the exchange tensors are kept, with \eref{eqn16b} ensuring that the total anisotropy energy between the out-of-plane and in-plane direction remains the same. Note that the tensor $\mathcal{K}^{\textrm{scaled}}$ may still be described by a single independent parameter $K^{zz,\textrm{scaled}}$ as required by symmetry.

The dependence of the spin wave modes of a skyrmion ($M=1$) and an antiskyrmion ($M=-1$) on the scaling parameter $\varepsilon$ is illustrated in figure~\ref{fig1}. As expected from the considerations above, for $\varepsilon=0$ modes with the same value of $mM$ are degenerate. It should also be noted that the frequency of the $m=0$ mode in this limit is zero, corresponding to the free choice of the helicity $\varphi_{0}$ in \eref{eqn13}. As the Dzyaloshinsky--Moriya interaction and the two-site anisotropies are introduced, the degeneracy of the modes between the skyrmion and the antiskyrmion is lifted, and more eigenmodes appear for the skyrmion than for the antiskyrmion below the excitation gap of the collinear background. 
%similarly to how the energy of the spin structures themselves becomes different. 
Since these types of interactions break the rotational freedom of the helicity, the $m=0$ mode evolves into the breathing mode of the skyrmionic structures, having been studied in many previous works~\cite{Schutte,Mochizuki}. In the antiskyrmion, the breathing mode remains the lowest-lying mode for $\varepsilon=1$, explaining the burst instability observed at negative field values (cf. table~\ref{table1}), while for the skyrmion the $m=-2$ elliptical mode assumes a lower frequency, leading to a strip-out instability as in Dzyaloshinsky--Moriya systems without the frustrated Heisenberg interactions~\cite{Bogdanov3,Schutte}.

%\subsection{Lattice discretisation effects\label{sec2c}}

%Since the rotational symmetry of the micromagnetic model in \eref{eqn8} is broken by the atomic positions, the assumptions about the cylindrically symmetric solutions in Eqs.~\eref{eqn15} and \eref{eqn16} no longer hold in this limit. The skyrmionic structures become distorted and their symmetry reduces to the $C_{3\textrm{v}}$ group of the lattice. For the structures with $M\neq 1$, which already became distorted in order to minimize the energy contribution from the Dzyaloshinsky--Moriya interaction, preferred orientations with respect to the underlying lattice can be observed~\cite{Rozsa2}. These orientations are robust against small perturbations such as weak driving currents~\cite{Kerber} or thermal fluctuations at low temperature~\cite{Weissenhofer}.

Due to the reduction of symmetry by the distorted spin structures and by the atomic lattice, the spin wave modes will no longer be eigenfunctions of the angular momentum of the form $\textrm{e}^{\textrm{i}m\varphi}$, instead they become linear combinations of them. To characterize the contribution of the different angular momentum eigenvalues $m$ to the lattice eigenmodes $\bi{r}^{k}$, the following quantities were introduced:
\begin{eqnarray}
\chi_{m}^{k,p}=\sum_{i}\textrm{e}^{-\textrm{i}m\varphi_{i}}S_{i}^{\perp,k,p},\label{eqn17}
\end{eqnarray}
where $i$ goes over the lattice sites and $p=x,y,z$ is a Cartesian component. $\bi{S}_{i}^{\perp,k}$ is obtained from $\bi{r}^{k}$ by rotating the eigenmodes back to the global coordinate system at each lattice site. The angle $\varphi_{i}$ is measured with respect to the center of the skyrmionic structure, which was chosen to be the center of mass of the out-of-plane spin components $S_{i}^{z}-1$, where $1$ is subtracted to remove the contribution of the out-of-plane collinear background.
%with the contribution of the collinear background subtracted.

If the breaking of the symmetry is weak, then $\chi_{m}^{k,p}$ will only take a significant value for a single $m$ quantum number for each mode $k$, and the angular momentum eigenvalues may still be used to approximately describe the excitations. This is the case for the $M=1$ skyrmion, where the Dzyaloshinsky--Moriya interaction does not distort the shape of the structure, and the reduction of the symmetry by the lattice is typically negligible if their size is considerably larger than the lattice constant. It was found in the simulations that the same remains a good approximation for the $M=-1$ antiskyrmion. The $m$ values shown in figure~\ref{fig1} represent the angular momentum eigenmode for which $\chi_{m}^{k,p}$ was maximal, which did not change for any of the modes as the parameter $\varepsilon$ was varied. 

\begin{figure}
\includegraphics[width=\columnwidth]{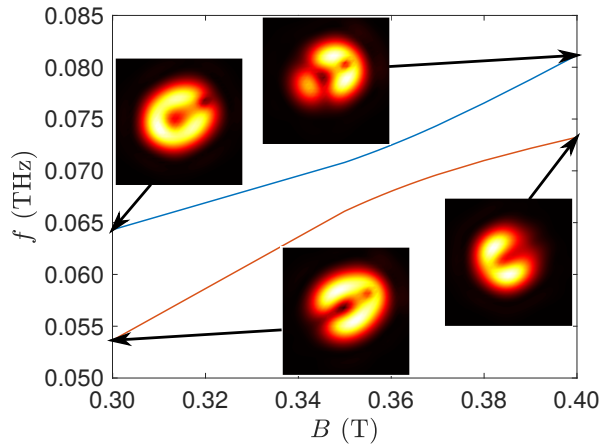}
\caption{Field dependence of two selected excitation modes of the chimera skyrmion. The insets illustrate the spatial distributions of the spin wave modes, via a contour plot of the squared amplitudes $\left|\bi{S}_{i}^{\perp,k}\right|^{2}$. Compare figure~\ref{fig2} for the real-space spin structure of the chimera skyrmion. The magnetic field was oriented outwards from the surface.\label{fig3}}
\end{figure}

This no longer remains true for more distorted spin structures. The angular momentum projections $\chi_{m}^{k,p}$ of a single mode $k$ may assume similar magnitudes for various $m$ values, and the dominant angular character may change as the parameters, such as the external magnetic field, are varied. A typical consequence of this effect is the hybridization between eigenmodes close in frequency, leading to avoided crossings in the spectrum. An example for this is shown in figure~\ref{fig3} for the chimera skyrmion. The frequency of two excitation modes is visualized in the figure as a function of external field. One of the modes is primarily localised on the lower left part of the chimera skyrmion with positive local vorticity (cf. figure~\ref{fig2} for the real-space configuration), while the other has a higher magnitude at the upper right part with negative local vorticity. Instead of the crossing of these two modes, their frequency difference remains on the order of $5\,\textrm{GHz}$ in the whole considered field regime, while the spatial character of the modes is gradually transformed as the magnetic field is changed. Avoided crossings do not occur in the continuum model between modes with different angular momentum quantum numbers, since they belong to different irreducible representations of the rotation operator and cannot hybridize. Even in calculations performed in the lattice model, the frequency difference between the modes may become so small that avoided crossings are not observable -- see, e.g., Ref.~\cite{Rozsa3}.

%In the continuum model, the skyrmionic structures always possess at least one zero-frequency spin wave mode, arising due to the translational invariance of \eref{eqn10}. Since continuous degeneracy is reduced to a discrete translational symmetry in the lattice model, the Goldstone modes of the continuum description acquire a small but finite frequency here. This frequency may become comparable in magnitude to the other excitations if the size of the skyrmionic structure is reduced to a few lattice constants~\cite{Bessarab}.

\subsection{Linear response\label{sec2d}}

In order to model the detection of the various excitation modes of the skyrmionic structures in experiments or numerical simulations, consider an excitation of the equilibrium magnetic structure by a small external magnetic field $\delta\bi{B}^{\textrm{ext}}$. Adding this excitation to the linearized Landau--Lifshitz--Gilbert equation \eref{eqn7}, one obtains
\begin{eqnarray}
\partial_{t}\tilde{\bi{S}}^{\perp}=\bi{D}_{\textrm{SW}}\tilde{\bi{S}}^{\perp}-\gamma'\left(-\textrm{i}\bi{\sigma}^{y}-\alpha\right)\delta\tilde{\bi{B}}^{\textrm{ext},\perp},\label{eqn18}
\end{eqnarray}
where $\delta\tilde{\bi{B}}^{\textrm{ext},\perp}$ is written in the local frame of reference and only its components perpendicular to the local spin direction are kept. A harmonic external excitation of frequency $\Omega$, $\delta\tilde{\bi{B}}^{\textrm{ext},\perp}\propto\textrm{e}^{-\textrm{i}\Omega t}$, drives the spin system into an oscillation with the same frequency after the decay of the transient eigenoscillations. The amplitude of this driven motion is given by the solution of the inhomogeneous algebraic equation
\begin{eqnarray}
\left(-\textrm{i}\Omega-\bi{D}_{\textrm{SW}}\right)\tilde{\bi{S}}^{\perp}=-\gamma'\left(-\textrm{i}\bi{\sigma}^{y}-\alpha\right)\delta\tilde{\bi{B}}^{\textrm{ext},\perp}.\label{eqn19}
\end{eqnarray}

Equation~\eref{eqn19} may be solved based on the spectral decomposition of the dynamical matrix,
\begin{eqnarray}
\bi{D}_{\textrm{SW}}=\sum_{k}-\textrm{i}\omega_{k}\bi{P}_{k}=\sum_{k}-\textrm{i}\omega_{k}\bi{r}^{k}\bi{l}_{k},\label{eqn20}
\end{eqnarray}
where the $\bi{l}_{k}$ are the left eigenvectors of $\bi{D}_{\textrm{SW}}$ with frequency $\omega_{k}$. Note that since $\bi{D}_{\textrm{SW}}$ is not a normal matrix (i.e., $\bi{D}_{\textrm{SW}}\bi{D}^{\dag}_{\textrm{SW}}\neq\bi{D}^{\dag}_{\textrm{SW}}\bi{D}_{\textrm{SW}}$), its left and right eigenvectors are not adjoints of each other. This implies that the projectors $\bi{P}_{k}$ will not be self-adjoint, but they will nevertheless satisfy
\begin{eqnarray}
\sum_{k}\bi{P}_{k}&=&\bi{I},\label{eqn21}
\\
\bi{P}_{k}\bi{P}_{k'}&=&\delta_{kk'}\bi{P}_{k}.\label{eqn22}
\end{eqnarray}
If certain eigenvalues coincide, $\bi{D}_{\textrm{SW}}$ may become non-diagonalisable, in which case \eref{eqn20} has to be extended by additional terms~\cite{Ashida}. Since the eigenvalues are always found in pairs of $\omega_{k}$ and $-\omega^{*}_{k}$~\cite{Rozsa3}, 
% due to the charge conjugation symmetry
this degeneracy typically arises for zero-frequency modes~\cite{Flynn}, which are present in skyrmionic structures in the continuum limit. However, the dynamical matrix becomes diagonalisable if these modes attain a finite frequency due to lattice discretisation,  %as discussed in section~\ref{sec2c}
or by including a finite Gilbert damping parameter $\alpha$.

Using \eref{eqn20}-\eref{eqn22}, the amplitude of the oscillation from \eref{eqn19} may be expressed as
\begin{eqnarray}
\tilde{\bi{S}}^{\perp}&=&-\gamma'\sum_{k}\frac{1}{-\textrm{i}\Omega+\textrm{i}\omega_{k}}\bi{P}_{k}\left(-\textrm{i}\bi{\sigma}^{y}-\alpha\right)\delta\tilde{\bi{B}}^{\textrm{ext},\perp}\nonumber
\\
&=&\bi{\chi}^{\perp}\left(\Omega\right)\delta\tilde{\bi{B}}^{\textrm{ext},\perp},\label{eqn23}
\end{eqnarray}
where the transversal susceptibility tensor $\bi{\chi}^{\perp}\left(\Omega\right)$ was introduced. The absorption or dissipated power is proportional to the imaginary part of the scalar product of the external field and the amplitude of the driven oscillation,
\begin{eqnarray}
P_{\textrm{diss}}&\propto&-\textrm{Im}\left(\left(\delta\tilde{\bi{B}}^{\textrm{ext},\perp}\right)^{\dag}\tilde{\bi{S}}^{\perp}\right)\nonumber
\\
&=&-\textrm{Im}\left(\left(\delta\tilde{\bi{B}}^{\textrm{ext},\perp}\right)^{\dag}\bi{\chi}^{\perp}\left(\Omega\right)\delta\tilde{\bi{B}}^{\textrm{ext},\perp}\right).\label{eqn24}
\end{eqnarray}

Equation \eref{eqn24} may be expressed in a different form in the limit of low damping. For $\alpha=0$, the left and right eigenvectors are connected by the relation
\begin{eqnarray}
\bi{l}_{k}=\left(\bi{\sigma}^{y}\textrm{sgn}\:\omega_{k}\bi{r}^{k}\right)^{\dag},\label{eqn25}
\end{eqnarray}
as can be proven by taking the adjoint of \eref{eqn7} and using that $\bi{H}_{\textrm{SW}}$ is a self-adjoint matrix. The sign term $\textrm{sgn}\:\omega_{k}$ is necessary to ensure the normalization $\bi{l}_{k}\bi{r}^{k'}=\delta_{k}^{k'}$, cf. \eref{eqn20} and \eref{eqn22}. Based on \eref{eqn25}, the right-hand side of \eref{eqn24} may be expressed as
\begin{eqnarray}
&&\left(\delta\tilde{\bi{B}}^{\textrm{ext},\perp}\right)^{\dag}\bi{\chi}^{\perp}\left(\Omega\right)\delta\tilde{\bi{B}}^{\textrm{ext},\perp}\nonumber
\\
&&=-\sum_{k}\frac{\gamma\:\textrm{sgn}\:\omega_{k}}{\Omega-\omega_{k}}\left|\left(\delta\tilde{\bi{B}}^{\textrm{ext},\perp}\right)^{\dag}\bi{r}^{k}\right|^{2}.\label{eqn26}
\end{eqnarray}
The scalar product $\left(\delta\tilde{\bi{B}}^{\textrm{ext},\perp}\right)^{\dag}\bi{r}^{k}$ may be evaluated in the global frame of reference. Since the typical size of the considered skyrmionic structures is on the order of a few nanometres, even for optical excitations it can be assumed that the external field is homogeneous in space, i.e., it belongs to the $m=0$ angular momentum eigenmode. If a linearly polarized field is applied along the $p=x,y,z$ direction, the response of the system is given by
\begin{eqnarray}
\left(\delta\tilde{\bi{B}}^{\textrm{ext},\perp}\right)^{\dag}\bi{r}^{k}=\chi_{m=0}^{k,p}B^{\textrm{ext}},\label{eqn27}
\end{eqnarray}
where $B^{\textrm{ext}}$ is the amplitude of the field and $\chi_{m}^{k,p}$ is introduced in \eref{eqn17}.

For finite $\alpha$, the leading-order correction is given by the appearance of an imaginary part of $\omega_{k}$ in the denominator in \eref{eqn26}. This gives rise to the dissipated power
\begin{eqnarray}
 P_{\textrm{diss}}\propto&&\left(B^{\textrm{ext}}\right)^{2}\sum_{\omega^{(0)}_{k}>0}\gamma\left|\chi_{m=0}^{k,p}\right|^{2}\alpha_{k,\textrm{eff}}\omega^{(0)}_{k}\nonumber
\\
&&\times\left\{\left[\left(\Omega-\omega^{(0)}_{k}\right)^{2}+\left(\alpha_{k,\textrm{eff}}\omega^{(0)}_{k}\right)^{2}\right]^{-1}\right.\nonumber
\\
&&\left.+\left[\left(\Omega+\omega^{(0)}_{k}\right)^{2}+\left(\alpha_{k,\textrm{eff}}\omega^{(0)}_{k}\right)^{2}\right]^{-1}\right\}.\label{eqn28}
\end{eqnarray}
Here $\omega^{(0)}_{k}$ denotes the excitation frequency for $\alpha=0$. $\alpha_{k,\textrm{eff}}=\textrm{Im}\:\omega_{k}/\textrm{Re}\:\omega_{k}\approx\textrm{Im}\:\omega_{k}/\omega^{(0)}_{k}$ is the effective damping parameter, which is mode-dependent and enhanced compared to the Gilbert damping~\cite{Rozsa3}. The summation is performed over eigenmodes with $\omega^{(0)}_{k}>0$, since the eigenvalues always appear in $\pm\omega^{(0)}_{k}$ pairs, which acquire the same imaginary part for finite $\alpha$. The negative frequencies are explicitly written out in the third line of \eref{eqn28}. The frequency dependence of $P_{\textrm{diss}}$ should be comparable to the absorption rate in spectroscopic measurements~\cite{Schwarze} or to the power spectral density calculated in numerical simulations~\cite{Kim,Lonsky}.

The contribution of the different eigenmodes $k$ to the dissipated power in \eref{eqn28} is determined by two factors. First, there is a frequency dependence in the form of a sum over two Lorentzian functions centred at $\pm\omega^{(0)}_{k}$. This naturally expresses that the eigenmodes may be excited most efficiently if the external frequency is in resonance with the eigenmode. However, this excitation is only possible if $\left|\chi_{m=0}^{k,p}\right|^{2}$ is sufficiently high, which is governed by selection rules. Since the external magnetic field couples to the magnetic dipole moment of the system, it can only excite modes in which this dipole moment is oscillating. The transversal deviations from the spin structure defined in spherical coordinates in \eref{eqn9a} may be written as
\begin{eqnarray}
\bi{S}^{\perp}=\left[\begin{array}{c}\cos\Theta_{0}\cos\Phi_{0}\delta\Theta-\sin\Theta_{0}\sin\Phi_{0}\delta\Phi \\ \cos\Theta_{0}\sin\Phi_{0}\delta\Theta+\cos\Theta_{0}\cos\Phi_{0}\delta\Phi \\ -\sin\Theta_{0}\delta\Theta\end{array}\right].\label{eqn29}
\end{eqnarray}
Assuming the cylindrically symmetric solutions in \eref{eqn15} and \eref{eqn16}, from the orthogonality of the angular momentum eigenmodes it is straightforward to obtain that modes satisfying $m\pm M=0$ may be excited by an in-plane field $p=x,y$, while modes with $m=0$ react to an out-of-plane field $p=z$. It is important to note that the type of the activated mode depends on the orientation of the field: although the external field is always homogeneous in the global frame of reference, the vorticity of the skyrmionic structure $M$ twists the in-plane field such a way that eigenmodes with a finite angular momentum quantum number $m$ may be detected. For the normal skyrmion with $M=1$, it has already been verified by numerical simulations~\cite{Mochizuki} and experiments~\cite{Schwarze,Satywali} that the homogeneous $m=0$ breathing modes may be excited by an out-of-plane field, while the $m=\pm 1$ gyration modes react to an in-plane field. Concerning skyrmionic structures with different topological charges, it is remarkable that the modes active to in-plane excitations depend on the absolute value of the vorticity, i.e., higher-charge skyrmions twist the in-plane field multiple times.

\begin{figure}
\includegraphics[width=\columnwidth]{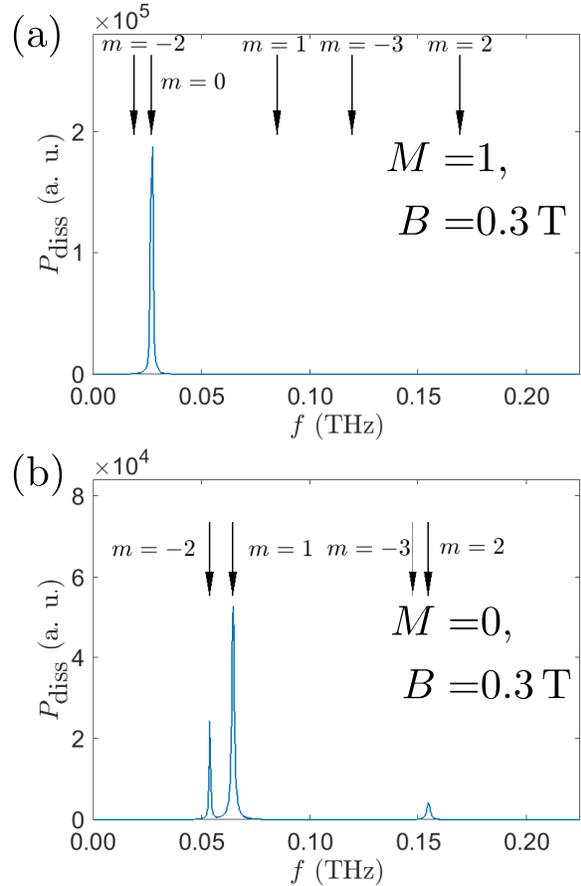}
\caption{Dissipated power $P_{\textrm{diss}}$ calculated based on \eref{eqn28} for (a) the $M=1$ skyrmion and (b) the chimera skyrmion, for an oscillating field applied along the $z$ direction.  The Gilbert damping was set to $\alpha=0.005$. Arrows denote the frequencies of the eigenmodes of the structure from \eref{eqn7}, the angular momentum quantum numbers $m$ indicate the value for which $\chi_{m}^{k,z}$ in \eref{eqn17} is maximal.\label{fig4}}
\end{figure}

Since the spin-wave modes of the distorted skyrmionic structures are described by a linear combination of angular momentum eigenmodes, these selection rules may no longer be strictly applied. More eigenmodes acquire a finite dipole moment and thus it becomes possible to excite them using an external magnetic field. This is illustrated in figure~\ref{fig4}. For the $M=1$ skyrmion in figure~\ref{fig4}(a), which is cylindrically symmetric apart from minor distortions due to the triangular lattice, only the $m=0$ breathing mode displays a considerable susceptibility to an oscillating field applied along the $z$ direction, as expected from the above argument. In contrast, three different spin wave modes of the chimera skyrmion react to the same type of excitation.
% Furthermore, since the competition between the Dzyaloshinsky--Moriya interaction and the frustrated Heisenberg interactions distorts the shape of the skyrmionic structures, the eigenmodes become hybridized, meaning that $\left|\chi_{k,m_{0}=0}^{p}\right|^{2}$ will remain finite for multiple excitation modes.

\begin{figure*}
\centering
\includegraphics[width=1.8\columnwidth]{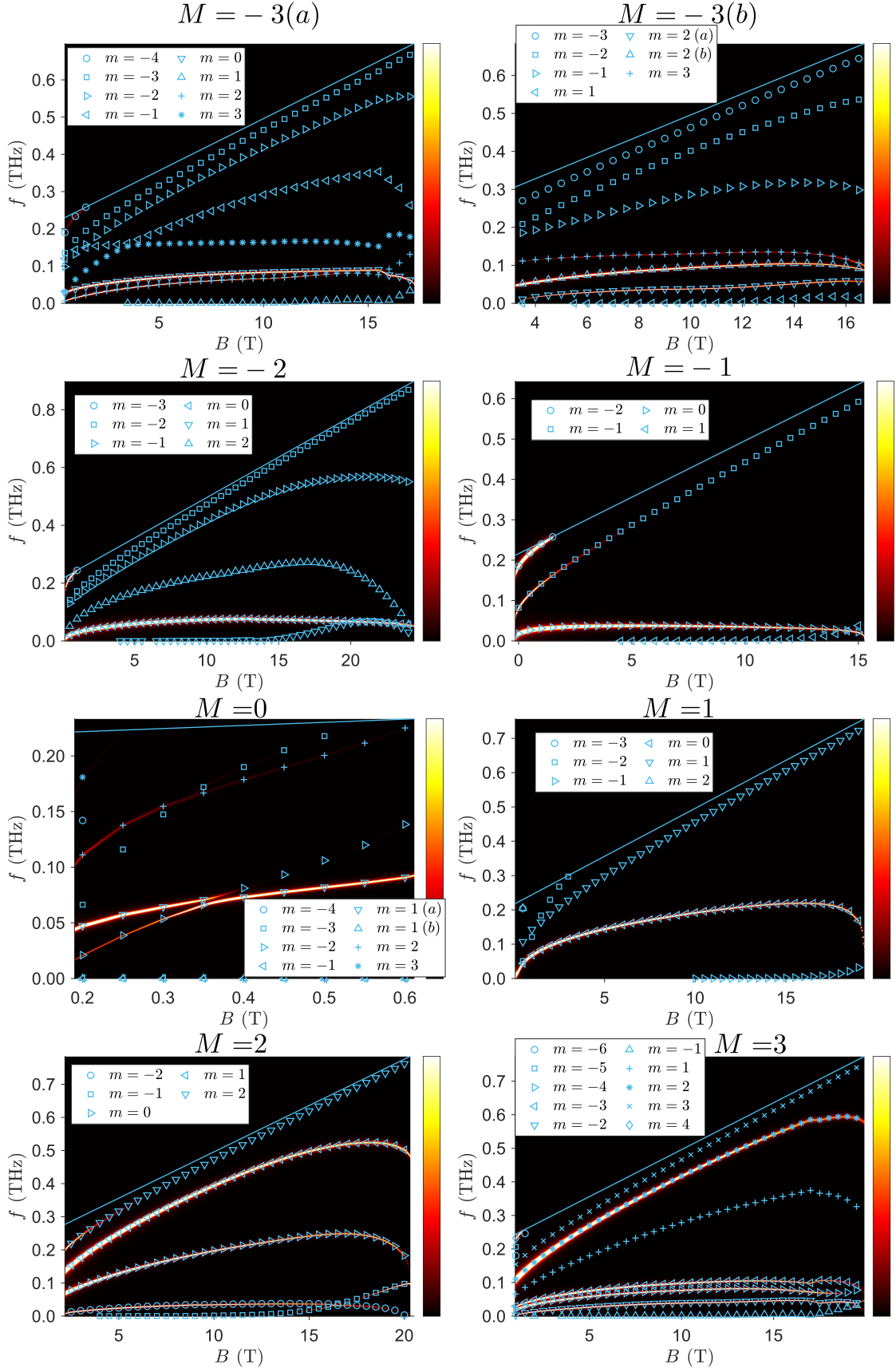}
\caption{Spin wave modes of the skyrmionic structures shown in figure~\ref{fig2}. Symbols display the frequencies of the eigenmodes based on \eref{eqn7}, the angular momentum quantum numbers $m$ indicate the value for which $\chi_{m}^{k,z}$ in \eref{eqn17} is maximal for the widest field range. Solid blue lines show the lowest-lying excitation of the collinear background. The background plot shows the dissipated power (see \eref{eqn28}) for an oscillating field applied along the $z$ direction, with black denoting low and bright denoting high intensities, respectively. The Gilbert damping for the calculation of the dissipated power was set to $\alpha=0.005$.\label{fig5}}
\end{figure*}

The calculated excitation frequencies of all skyrmionic structures from figure~\ref{fig2} in the whole range of their stability are summarized in figure~\ref{fig5}. Only the localised eigenmodes below the continuum spectrum of the collinear background can be identified~\cite{Schutte,Rozsa2}, the lower bound of which is denoted by continuous lines. Note that this lower bound is below the ferromagnetic resonance frequency belonging to the spatially homogeneous excitation mode, since the minimum of the spin wave spectrum is located close to the minimum of the Fourier transform of the Heisenberg exchange interactions $J\left(\bi{q}\right)$, which is found at finite wave vector as shown in \cite{Rozsa}.

The dissipated power $P_{\textrm{diss}}$ calculated from \eref{eqn28} is plotted in color in the background, with the excitation field oriented along the out-of-plane direction. Therefore, the bright lines indicate modes reacting strongly to the external excitation. Apart from the almost perfectly cylindrically symmetric $M=1$ skyrmion, all structures display at least two modes which could be detected by out-of-plane probing fields. The shift of the intensity from one spin wave branch to the other close to the avoided crossings indicates the hybridization of these modes, which is especially apparent for the chimera skyrmion also shown in figure~\ref{fig3}. The spin wave modes for which the susceptibility to the out-of-plane fields is negligible may often be excited by in-plane fields instead, which are sensitive to the $m=\pm M$ angular momentum components of the modes. Note that the different symbols used for the various eigenmodes and the assigned $m$ quantum numbers are only meant to be a guide to the eye, since apart from the $M=1$ skyrmion the angular character of the modes gradually changes with the field due to the hybridization.

The lowest-lying eigenmodes are located very close to zero at low field values. In the continuum model, these correspond to Goldstone modes arising due to the translational invariance of \eref{eqn10}. For the skyrmions this is the $m=-1$ mode for the magnetic field oriented out-of-plane, while for the antiskyrmions it is the $m=1$ mode due to the symmetry discussed in section~\ref{sec2b2}. The chimera skyrmion possesses two Goldstone modes $m=\pm1$. This can be explained by its vanishing net topological charge, similarly to the so-called $2\pi$ skyrmion in \cite{Rozsa3} or to antiferromagnetic skyrmions in \cite{Kravchuk2}. Since this continuous degeneracy is reduced to a discrete translational symmetry in the lattice model, these modes 
% of the continuum description 
acquire a small but finite frequency here. At high field values the size of the skyrmionic structures is reduced to a few lattice constants, and this frequency becomes comparable in magnitude to that of the other excitations~\cite{Bessarab}. The lattice discretisation effect is also responsible for the unusual behaviour of the eigenfrequencies in the $M=-3\,(a)$ configuration close to the upper critical field: the center of the structure shifts from a lattice site to the hollow position in the middle of four neighbouring lattice sites at around $B=16\,\textrm{T}$ where the discontinuities occur.

\section{Conclusion\label{sec4}}

In summary, the shape and stability as a function of magnetic field of skyrmionic spin structures with various topological charges was investigated, using the magnetic interaction parameters determined from \textit{ab initio} calculations for the (Pt$_{0.95}$Ir$_{0.05}$)/Fe/Pd(111) system~\cite{Rozsa}. It was shown that the Dzyaloshinsky--Moriya interaction and two-site compass-like anisotropy terms distort the structures from the cylindrically symmetric shape observed for only frustrated Heisenberg interactions and on-site anisotropy, and select preferred values for the helicity. For the antiskyrmion with vorticity $M=-3$, objects with two different helicities were found to coexist in a wide range of external fields, caused by the incompatibility between the intrinsic $C_{4}$ rotational symmetry of the $M=-3$ structure and the $C_{3\textrm{v}}$ group of the lattice. Although the $M=1$ skyrmion is energetically preferred by the Dzyaloshinsky--Moriya interaction, it was demonstrated that the $M=-1$ antiskyrmion remains stable at lower and the $M=-2,2,3$ structures up to higher field values.

Concerning the localised spin wave modes of the textures, it was shown that in the absence of distortion-inducing interactions, the excitation modes of skyrmions and antiskyrmions with opposite vorticities $M$ have the same frequencies, if the angular momentum $m$ of the spin-wave mode is also reversed. This degeneracy is lifted as the skyrmionic structures become distorted, and the $m=0$ Goldstone mode belonging to the freedom of the helicity evolves into a breathing mode with finite frequency. In the not cylindrically symmetric structures, the excitation modes are no longer described by a single angular momentum quantum number, leading to hybridizations between them and avoided crossings in the spectrum as the external field is changed. It was demonstrated that multiple modes may be excited in each distorted skyrmionic structure by oscillatory external fields, while in the cylindrically symmetric structures the detection of the possible excitations is restricted by the selection rules to $m=0$ for out-of-plane fields and $m=\pm M$ for in-plane fields.

These results should motivate experimental and theoretical studies aimed at the detection of diverse skyrmionic structures and their application for magnon-based computing. Since the spin configurations investigated here remain stable up to magnetic fields on the order of $10\,\textrm{T}$, their excitation frequencies reach hundreds of GHz. Based on the linear scaling between the magnetic field and the spin wave frequencies (e.g., $\omega_{\textrm{Larmor}}=\gamma B$ for noninteracting spins), for skyrmions stabilised by fields in the range of $100\,\textrm{mT}$, the spin wave frequencies are of the order of a few GHz, as has been demonstrated in experimental studies on various systems~\cite{Schwarze,Onose,Satywali}. The guidelines derived for the hybridization of the angular momentum eigenmodes and their response to an oscillating external field are expected to be relevant for other localised spin structures lacking cylindrical symmetry, regardless of their size or the field strength required to stabilise them. These may be formed by applying mechanical strain externally~\cite{Shibata,Sukhanov}, due to the lattice mismatch~\cite{Hsu} or intermixing~\cite{Meyer} in interfacial systems, and in skyrmion lattices~\cite{Ikka,Koide} or isolated skyrmions, antiskyrmions~\cite{Kuchkin} and bimerons~\cite{Gobel} stabilised by oblique or in-plane fields. The hybridization between modes with different angular momentum quantum numbers in distorted skyrmionic structures may enable the generation of spin waves with a finite orbital angular momentum~\cite{Jia} by exciting the system with a homogeneous external field.

%These include mechanical strain applied externally \cite{Shibata,Sukhanov} or appearing due to the lattice mismatch in interfacial systems \cite{Hsu}, noncylindrically symmetric skyrmions in the conical phase of bulk chiral magnets \cite{Leonov}, and skyrmions stabilised in ultrathin films by a tilted external magnetic field \cite{Ikka,Koide}. Furthermore, there have been indications that skyrmions may be formed embedded in tilted cycloidal spin spiral phases in two-dimensional systems. In particular, it was demonstrated by numerical simulations in Ref.~\cite{Ikka} that besides the previously observed clockwise and counterclockwise gyration modes and the breathing mode of the skyrmion lattice stabilised by the Dzyaloshinsky--Moriya interaction, an additional excitation mode may be observed when the stabilizing field is not perpendicular to the surface. This additional mode was classified as an additional clockwise gyration mode based on the investigation of its spatial profile. The hybridization between modes with different angular momentum quantum numbers in distorted skyrmionic structures may enable the generation of spin waves with a finite orbital angular momentum \cite{Jia} by exciting the system with a homogeneous external field.

\ack
Financial support of the Alexander von Humboldt Foundation, the National Research, Development and Innovation Office of Hungary via Project No. K131938 and the German Research Foundation via Project No. 403502522 is gratefully acknowledged.

\end{document}